\documentstyle[pra,aps,twocolumn,epsf]{revtex}
\begin{document}
\draft

\title{EXPERIMENTAL INVESTIGATIONS OF SPATIAL DISTRIBUTION ANISOTROPY
OF PULSED PLASMA GENERATOR RADIATION.}
\vskip15pt

%{\it
\author{Yu.A.Baurov\thanks{baurov@www.com}*, I.B.Timofeev**, V.A.Chernikov**, S.F.Chalkin**}

\address{* Central Research Institute of Machine Building,  141070, Pionerskaya 4, Korolyov, Moscow region.}

\address{** Moscow State University named by M.V.Lomonosov, Department of Physics, Chair of Physical Electronics,
119899, Vorob'evy Gory, Moscow.}
\date{\today}

\maketitle
\begin{abstract}
Results of experimental investigation of plasma luminous
emittance (integrated with respect to time and quartz
transmission band spectrum) of a pulsed plasma generator
depending on its axis spatial position, are presented. It is
shown that the spatial distribution of plasma radiant intensity
is of clearly anisotropic character, that is, there exists a
cone of the plasma generator axial directions in which the
radiation of plasma reaches its peak. A possible explanation of
the results obtained is given based on a hypothesis of global
anisotropy of space caused by the existence of a cosmological
vectorial potential $\bf{A}_g$. It is shown that the vector $\bf{A}_g$ has the
following coordinates in the second equatorial coordinate
system: right ascension $\alpha = 293^\circ \pm 10^\circ$,
declination $\delta = 36^\circ \pm 10^\circ$.
The experimental results are in accordance with those of
the earlier experiments on determining the direction of $\bf{A}_g$.
\end{abstract}

\section{Introduction.}

In Refs. \cite{1,2,3,4,5,6,7,8,9,10,11,12}, a new assumed interaction of objects in nature
distinct from the four known ones (the strong, weak,
electromagnetic, and gravitational interactions) is predicted
and investigated.  The new force is caused by the existence of
the cosmological vectorial potential $\bf{A}_g$, a new fundamental
vectorial constant entering into the definition of discrete
objects, byuons. According to the hypothesis advanced in Refs.
\cite{1,2,3,4}, in the process of minimization of potential energy of
interaction between byuons in the one-dimensional space formed
by them, the observable physical space as well as the world of
elementary particles together with their properties appear. The
masses of particles in the model proposed are proportional to
the modulus of the summary potential $\bf{A}_{\Sigma}$  which contains $\bf{A}_g$ and
vectorial potentials of various magnetic sources as of natural
origin (from the Earth, the Sun, etc.) so of artificial origin
(for example, the vectorial potential $\bf{A}$ of magnetic fields from
solenoids, plasma generators, etc.). The value $|\bf{A}_{\Sigma}|$ is always
lesser than $|\bf{A}_g| \approx 1,95\times10^{11} Gs\cdot cm$ \cite{1,2,3,4,5,6,7}.
The vectors $\bf{A}_{\Sigma}$ and $\bf{A}_g$
are practically always collinear because of the great value of
the latter.  In the model of Refs. \cite{1,2,3,4}, the process of
formation of the physical space and charge numbers of elementary
particles is investigated. Therefore, in contrast with
calibration theories (for example, the classical and quantum
field theories), the values of potentials acquire the physical
sense, which is in tune with the known and experimentally tested
Aharonoff-Bohm effect \cite{13,14,15,16} as a particular case of quantum
properties of space described in Refs. \cite{1,2,3,4}.

The on-earth experiments (with high-current magnets \cite{1,2,5,6,7}, with a
gravimeter and an attached magnet \cite{1,2,8}), investigations of
changes in $\beta$-decay rate of radioactive elements under the action
of the new force \cite{9}, and astrophysical observations \cite{10,11}
have given the following approximate coordinates (in the second
equatorial system) for the direction of the vector  $\bf{A}_g$: right
ascension $\alpha \approx 270^\circ$, declination $\beta \approx 34^\circ$.

In the aggregate, the experiments carried out \cite{1,2,4,5,6,7,8,9,10,11,12} have shown
that if the vectorial potential of some current system is opposite
in direction to the vector $\bf{A}_{\Sigma}$ then the new force repels any substance out of the region of
weakened $|\bf{A}_{\Sigma}|$ mainly in the direction of $\bf{A}_g$. The magnitude of
the new force $\bf{F}$ in the experiments with high-current magnets
(magnetic flux $B$ being up to $15T$) was equal to $\sim 0,01-0,08g$
for the test body mass $\sim 30g$.  When investigating the new
interaction with the aid of a stationary linear arc plasma
generator (with $\sim 60kW$, current $\sim 300A$, voltage $220V$, mass flow
rate $V\approx120ms^{-1}$) positioned on a special rotatable base, there
were detected two special directions corresponding to energy
release in the plasma jet up to $40\%$ more than the average energy in
the plasma flow during rotation of the plasma generator in the
horizontal plane through nearly $360^\circ$, with the summary
experimental error of $\pm12\%$. These directions laid left and right
from the vector $\bf{A}_g$ at an angle of $\sim 45^\circ-50^\circ$ with the latter, they
corresponded with the most efficient angle between the vector $\bf{A}$
of the current system and the vector $\bf{A}_g$ (i.e. with the maximum
magnitude of the new force) being equal to $135^\circ\div140^\circ$ \cite{12}.
The found directions of maximum action of the new force in the
experiments with the plasma generator gave the following  $\bf{A}_g$ -
coordinates: $\alpha \approx (280 - 297)^\circ$, $\delta \approx 30^\circ$.

The aim of the present work is further experimental investigation of the global space
anisotropy associated with $\bf{A}_g$.

\section{Formulation of the problem.}

The new force predicted in the Refs.\cite{1,2,3,4,5,6,7,8,9,10,11,12} is of complex nonlinear and
nonlocal character and can be represented in the form of some
series in $\Delta A$, a difference between changes in $\bf{A}_\Sigma$ at the location
points of a sensor and test body. For the first approximation of
that series we have
$$F\sim N\Delta A\frac{\partial\Delta A}{\partial x}$$
where $x$ is the spatial coordinate,
and $N$ is the number of stable elementary particles (electrons,
protons, neutrons) in a space region with $\bf{A}_\Sigma$ varying due to the
vectorial potential of some current system.  It was shown in the
experiments\cite{1,2} with rotating magnetic discs and an
engine-generator as well as in the experiments with the plasma
generator\cite{12} that the force $F$ can be substantially increased
(tens and more times) when phasing the motion of the body with
the process of physical space formation from byuons (i.e. the
working body must change $\bf{A}_\Sigma$ by its own potential $A$ and move in
the direction of $\bf{A}_g$. Therewith the particles of the body must
rotate in phase with the above-mentioned process of formation of
the physical space). In such a case energy will be taken from
the physical space through the elementary particles of the
working body. The law of energy conservation in the system
"working body - physical space" will be valid. As is known\cite{17},
the basic energy of the Universe ($> 90\%$) is determined by the
"dark" (virtual) matter. The model of formation of the physical
space\cite{1,2,3,4} describes the phenomenon of the {\it "dark matter"}
reasonably well.

Based upon the physics of the new assumed
force and mechanisms of strengthening it to realize the aims of
the present paper, the experimental installation should met the
following requirements.  First, it should realize a maximum $\Delta A$
corresponding to maximum possible values of current.  Second, to
realize a maximum $\frac{\partial\Delta A}{\partial x}$,
the current density should be as high as possible.
  Third, if a plasma generator is chosen for
investigation of the new force, its discharge should be
maintained in a medium with the most great value of $N$ (for
example, not in vacuum but in air at the atmospheric pressure or
in water).  Fourth, to realize the mechanism of strengthening
the new force (i.e. phasing the motion of the working body with
the process of space formation at the rate on the order of the
light speed), the magnitude of the velocity $V$ in the discharge
should be the maximum possible.  Fifth, the experimental
installation on the base of a plasma generator chosen for
investigating the global space anisotropy by way of scanning the
celestial sphere should introduce into the experiment minimum
systematic errors connected with the rotation of the plasma
generator in space (for example, with an influence of curvature
of hoses delivering water, air, and argon to a stationary plasma
generator, on heat release in its jet as in Ref.\cite{12}, or with
action of the Coriolis force on the flow of water in a measuring
tube\cite{12}, etc.).  All these requirements are most closely met
by the pulsed plasma generator (magnetoplasma compressor).

\section{Experimental installation and technique.}

The experimental installation was comprised of a pulsed plasma generator and a
system of measuring the plasma radiation. The plasma generator
was fed from an energy-storage capacitor $100mF$ in total capacity
with operating voltage up to $5kV$. The total energy accumulated
in the capacitor was equal to $\sim 1,25 kJ$. The battery was charged
from a standard high-voltage power source {\bf GOR-100} and commutated
to the load (pulsed plasma generator) with the aid of a trigatron
type air spark gape activated by a short ($\sim 1ns$) high-voltage
($\sim 30kV$) pulse coming to the air-gap from a trigger circuit.

 The design of the pulsed plasma generator (1) is shown in Fig.1. The
case (2) of the generator (its outer electrode being anode) was
made of thin-walled copper tube 11 mm in external diameter and
$100mm$ in length. The axial electrode (3) (cathode) $4mm$ in
diameter made from copper bar was placed into an acrylic plastic
tube (4) with inner diameter of $4mm$ and outside diameter equal
to that of the outer electrode. In the Fig.1 shown is also a
statistic average pattern (5) of discharge currents of the
plasma generator. The angle $\phi$ was equal to $\sim 30^\circ$. The whole
construction as a unit was an analogue of the coaxial plasma
accelerator.  The plasma generator was locked on a textolite
plate (6) positioned on a special adjustment table (8) rotatable
around its vertical axis (7). The table (8) was provided with a
limb (9) allowing to control the angle of rotation of the whole
system relative to some starting position. The plate (6) itself
could rotate around the horizontal axis (10) through an
arbitrary angle $\beta$. The system as an assembly made it possible to
rotate the plasma generator during the experiment around the
vertical axis (7) through any angle, and around the horizontal
axis through any angle in the range $-90^\circ<\beta<60^\circ$.
It was assumed that at $\beta<0$ the discharge of the plasma generator turns to the
surface of the Earth. The horizontal position of the plasma
generator corresponded to the angle $\beta=0$. The trajectory of motion of
the face of plasma generator during its rotation in the
horizontal plane represented a circle.  All experiments were
carried out in air at atmospheric pressure.

The volt-ampere characteristics measured with the aid of Rogovsky belt and an
induction-free voltage divider assembled from resistors of
{\bf TVO}-type, made it possible to judge the amount of energy put
into the discharge channel. Some typical volt-ampere
characteristics are shown in Fig.2.  One can see from the Figure
that the discharge was of quasi-periodic character but with
great damping. The quasi-period of the discharge current in
conditions of the experiment was equal to $\sim 70ms$. The amplitude
of the current in the first maximum reached $21kA$. The maximum
voltage between the electrodes equaled $3.5kV$ at $5kV$ charging
voltage on the energy-storage capacitor.  The dynamics of plasma
outflow (11 in Fig.1) as well as the characteristic dimensions
of plasma jet were investigated with the aid of a
super-high-speed photorecorder of {\bf SFR}-type operating in
single-frame filming mode. A fragment of the record is given in
Fig.3. The earlier studies of the plasma generator in use have
shown that about $(30-40)\%$ of its power were released in the optical
frequency range\cite{18}.  The absolute value of radiation energy
in quartz transmission band ($\lambda > 220nm$) was measured by a
thermal detector of {\bf LETI}-type (12 in Fig.1) rigidly fixed on the
plate (6 in Fig.1) so that the relative positions of the plasma
generator and thermal detector could not change as the plasma
generator rotated. In so doing, the axis of the thermodetector
was directed to the prior known range of maximum discharge glow
lying on the axis of the plasma generator $\sim 2cm$ from its face.

The thermodetector was calibrated with the use of a standard
radiation source of {\bf IFP-1200} type giving $E_{st}=35.64J/sr$ of
radiant energy per unit of solid angle. In the process of
calibrating and measuring, the signal from the thermodetector
came to the input of a mirror-galvanometer oscillograph {\bf K117}.
The radiant energy of plasma was calculated from the formula
$$ E = 4\pi A l^2 n,$$
where $A$ is the calibration coefficient, $l$ is the
distance from the radiation source (in meters), $n$ is the maximum
magnitude of signal on the strip of oscillograph (in
millimeters).  At the characteristic dimensions of plasma
cumulation zone of the order of $1cm$ and the distance from the
thermodetector to the plasma source $l = 20cm,$ the radiation
source could be taken as a point one. Because the radiative
energy of plasma is proportional to $T^4$, an insignificant change
in temperature $T$ could be sensed by the thermodetector. Despite
a significant increase in discharge current and voltage of the
plasma generator considered in comparison with those in the
experiment of Ref.\cite{12}, the thermal effect of the new force
action was expected at a level of $10\%$ owing to short duration of
the discharge.  The main parameter measured in the experiment
 was the deflection of the beam of the mirror- galvanometer
oscillograph {\bf K117}. This deflection, proportional to the
radiative intersity of plasma, was recorded on the photographic
strip and gave information on the amount of energy released in
the discharge of the plasma generator and, while scanning the
celestial sphere, on the direction of maximum action of the new
force upon the particles of the plasma discharge.

  To investigate the direction of the global anisotropy of physical
space caused by the vector $\bf{A}_g$, as well as the new interaction
connected with this vector, the following technique was used. In
the first of experiments (15.12.1999-3.05.2000), the plasma
generator was rotated only around the vertical axis (7 in
Fig.1).

The start time of the experiment was determined by the
position of the vector $\bf{A}_g$ near the horizontal plane. On the
basis of previous experiments\cite{1,2,4,5,6,7,8,9,10,11,12}, the vector $\bf{A}_g$ was
assumed to have the following approximate coordinates:
$270^\circ<\alpha<300^\circ, 20^\circ<\delta<40^\circ$. The statistic average pattern of
discharge currents shown in Fig.1 was assumed to correspond to
direction of the maximum current along the axis of the plasma
generator. Therefore one could expect that the direction of axis
of the plasma generator allowed to judge the efficient angle of
action of the new force, i.e. the efficient angle between the
main discharge current and the vector $\bf{A}_g$. Recall that the new
force acts when the vector of the discharge current has a
component directed oppositely to $\bf{A}_g$, and the vector of
velocity $V$ of the particles points in $\bf{A}_g$ direction.  In the
experiments of 15.12.1999 and 20.01.2000, the luminous emittance
of plasma jet was measured by the thermodetector {\bf LETI} in 30
degree intervals through one complete revolution of the
adjustment table. In all other experiments the measurements of
the luminous emittance were made in 10 degree intervals.  In the
second sequence of experiments, to improve the direction of the
new force in space, the celestial sphere was scanned in the
vicinity of the extremum directions found in the previous
experiments during rotation of the plasma generator in the
horizontal plane.

\section{Results of experiments and discussion.}

From 15.12.1999 till 3.05.2000, 32 experiments with rotation of
the plasma generator around the vertical axis through $360^\circ$ were
carried out. The duration of one experiment (25 or 30 shots)
was no more than 30-25 min. As an illustration, in Fig.4 shown
are the values of deflection $L$ of the beam of the
mirror-galvanometer oscillograph (in millimeters) in dependence
on the angle of rotation $\theta$ in the experiments of 15.12.1999 and
20.01.2000.

 In the experiment of 15.12.1999 conducted from
$16^{40}$ till $17^{15}$, a burst in luminous emittance of plasma
generator discharge was observed at $16^{55}$. Therewith the angle $\theta$
measured from an arbitrary direction H shown in Fig.5 was equal
to $165^\circ$, and the value of emittance was $25.7\%$ above its average
over the duration of the experiment with a root-mean-square
error of $\pm 3.7\%$. In the experiment of 20.01.2000 performed from
$15^{35}$ till $16^{10}$, the burst of plasma luminous emittance was
detected at $\theta = 135^\circ$ (measured from that same direction H in
Fig.5) with $24\%$ excess over the average value of emittance at a
root-mean-square error of $\pm 3.3\%$.

For qualitative understanding of
the result obtained and the processing procedure, in Fig.5 are
shown by arrows some projections of the plasma generator axis on
the ecliptic plane corresponding to maximum luminous emittances
of plasma discharge with indication of concrete positions of the
Earth in the process of its orbiting around the Sun, the data of
the experiments, and the points in time at which the said
maximum (with values greater than the experimental error) were
observed. The dotted line denotes secondary extremum directions
for the emittance. In all experiments the plasma generator
rotated counter-clockwise if the plane of rotation seen from
above. The direction of the axis of the plasma generator at $\theta =0$
is indicated by letter H for each experiment.

At the center of Fig.5 (at the site of the Sun) a circle diagram summarizing
the results all experiments, is given. They were processed in
the following manner. The circle was divided in ten-degree
sectors so that the radius-vector passing from the center of the
circle along the initial boundary of the first sector was aimed
at the point of vernal equinox (21.03) from which the angular
coordinate $\alpha$ is counted anticlockwise in the second equatorial
system.

In Fig.5 the heights of crosshatched triangles with a
$20^\circ$-angle at the center of the circle are proportional to the
sums (in percentage) of extremum deflections of the oscillograph
beam from its average coordinate which stand out above the standard
error of measurements and fall within one or the other of the
triangles, for all bursts in luminous emittance observed in all
32 experiments. As is seen from Fig.5, the maximum emittances
were observed (more often and with maximum amplitudes) in the
$25^{th}, 26^{th}$ and $33^{th}, 34^{th}$ sectors.

Notice that when the vectorial potential of the plasma generator is directed exactly
opposite to $\bf{A}_g$, the change in $\bf{A}_\Sigma$ should be maximum and, hence, the
magnitude of the new force should be zero since $\frac{\partial\Delta A}{\partial x} = 0$.
It follows herefrom that the direction of the vector $\bf{A}_g$ must be related
with the sectors 29,30 (Fig.5). This direction has the
coordinates: $\alpha=290^\circ\pm10^\circ$, and an efficient angle between the
axial current of the plasma generator and the vector $\bf{A}_g$ being
equal to $140^\circ\pm10^\circ$.  The result obtained fully coincides (with an
error above indicated) with that of Ref.\cite{12} in which a
stationary plasma generator positioned on a special rotatable
base and a copper measuring tube with water passed through as a
sensing element located in the plasma jet, were used. The
results of the present paper do not contradict those of earlier
measurements of vector $\bf{A}_g$\cite{1,2,4,5,6,7,8,9} and are much more precise.  It
should be noted that a considerable number of bursts fall into
the sectors 11 and 12, corresponding to the direction precisely
opposite to that of $\bf{A}_g$. This may be attributed to the action of
side currents in the discharge of the plasma generator (see
Fig.1) directed at an angle $\phi \approx 30^\circ$ to its axis. That is, in
this case the side currents make an angle of $\sim 150^\circ$ with $\bf{A}_g$ which
is near to the most efficient angle of  $\sim 140^\circ$ found from the
direction of the main axial current of the plasma generator.
Therewith the vector of mass velocity {\bf V} of the discharge
particles is in opposition to $\bf{A}_g$.  Hence, the mechanism of
strengthening the new force is here ineffective as compared with
the situations in which the axis of the plasma generator fall
into the sectors 25,26 and 33,34 where the vectors {\bf V} and $\bf{A}_g$ are
directed to the same side.

From 10.05.2000 till 31.05.2000 and from 11.10.2000 till 3.11.2000,
a run of experiments was carried
out with scanning the celestial sphere in the vicinity of
sectors from $25^{th}$ till $34^{th}$ for determining most efficient angles of
special position of the plasma generator axis relative to the
vector $\bf{A}_g$, i.e.  the angles of maximum action of the new force.

As an illustration, in Table 1 the results of the last
experiment performed 3.11.2000 are presented. In all, 77 shots
were made with the average duration of scanning the celestial
sphere about 90min. The angle $\gamma$ in Table 1 corresponds to
rotation of the plasma generator around the vertical axis, and
the angle $\beta$ does to its rotation around the horizontal axis. In
each square of the Table, the magnitude of deflection of the
beam of the mirror-galvanometer oscillograph proportional to the
luminous emittance of the plasma discharge, is shown. In its
turn, this emittance is proportional to the value of the new
force acting on electrons and other particles of the discharge.

As the duration of the experiment was more than 1.5 hours  (the
rotation of the Earth through approximately $23^\circ$), the average
deflections ($L_{av}$) of the oscillograph beam were calculated for
one passage of the plasma generator through the angle $\gamma$ at the
angle $\beta$ fixed. The value $L_{av}$ and root-mean-square deflections $\sigma$
(in percentage) are shown in Table 1 for each passage of the
plasma generator. The start time of the experiment was chosen
from an expected time of fall of $\bf{A}_g$ into the range of horizontal
plane. In the experiment considered, the angles $\gamma = 220^\circ$ and
$\beta = -20^\circ$  correspond to a coincidence of the north-direction at the
place of the installation (Moscow Lomonosov University) with the
direction of the $\bf{A}_g$ - projection on the horizontal plane.

A summary result of vernal and autumnal experiments with scanning
the celestial sphere is shown in Fig.6. In this Figure given are
the relative spatial coordinates and directions (arrows) of only
those positions of the axis of the plasma generator at which the
deflection of the oscillograph beam in the process of discharge
was above the root-mean-square one. In Table 1 they are
asterisked for the experiment of 3.11.2000. The results of the
vernal ran of experiments are related to 11.25 of Moscow time of
31.05.2000 and noted by circles in Fig.6. The autumnal results
(also asterisked) are related to 20.00 of Moscow time of
11.10.2000.

When scanning the celestial sphere, the plasma
generator was positioned during the
experiment (at various angles $\beta$) on circles of different radii but with a deflection
from the level of some horizontal plane no more than $\pm12,5\%$ in the range of
angles $-20^\circ<\beta<40^\circ$. Therefore, for clearness, all positions of the
plasma generator are given in projection onto the horizontal
plane as for vernal so for autumnal runs of experiments with
indication of angles of the slope of the generator axis to this
plane (angle $\beta$). This is understandable since the sections of
cone by the plane are well known.

The ranges of an $\gamma$-angles
$170^\circ-190^\circ$ and $250^\circ-270^\circ$ in Fig.6 correspond in space to the
sectors 34,33 and 26,25 in Fig.5, respectively. The horizontal
planes for these data and the indicated related times of
experiments (31.05.2000, 11.25 and 11.10.2000, 20.00) intersect
at an angle  of $\sim 35^\circ$.

As is seen from Fig.6, with the
exception of one and only point ($\gamma = 25^\circ, \beta = -40^\circ$),
all other positions of the axis of the plasma generator in the vernal run of
experiments form a section of a cone of directions.  The
dispersion of points in the autumnal run of experiments is more
great but all experimental points except two ($\gamma = 220^\circ,
\beta = +10^\circ; \gamma = 225^\circ, \beta = +10^\circ$) fit into the cone shown in Fig.6 with
a dispersion of coordinates of axial directions equal to $\pm10^\circ$.

With that same error, the axial direction of the cone makes it
possible to find the direction of the vector $\bf{A}_g$ (since   along this
direction $\frac{\partial\Delta A}{\partial x}=0$) from the vectorial potential of the axial current
of the plasma generator and, hence, the new force should be
zeroth, too, i.e. bursts should be absent along $\bf{A}_g$.  This
direction is also shown in Fig.6. It has the coordinates
$\alpha \approx 293^\circ\pm10^\circ$ and $\delta \approx 36^\circ\pm10^\circ$
in the second equatorial system. The
direction indicated qualitatively coincides with the results of
earlier experiments\cite{1,2,4,5,6,7,8,9} and specifically with the recent
experiments carried out on the basis of stationary plasma
generator\cite{12} as well as with astrophysical observation of
anisotropy of distribution of solar flares and galactic pulsars
\cite{1,2,10,11}.

\section{Analysis of experimental errors.}

The errors in determining the direction of $\bf{A}_g$ can be classified
as systematic ($\sigma_{syst}$) and random, or statistic ($\sigma_{st}$) ones.
In the experiment in consideration, the systematic errors can be
due to the following causes:  initial bursts at the beginning of
experiment with rotation of the plasma generator associated with
deficient prior "warming up" the system (i.e. without a previous
"ranging fire" before the readings of the oscillograph flatten
out); a build up of the cathode of the plasma generator in the
course of the experiment giving rise to a change in geometry of
discharge currents (see Fig.1); burn-out of an isolator also
changing the pattern of currents in the discharge; a glass
turbidity of the thermodetector LETI; a withdrawal of its axis
from the direction of the maximum luminous emittance of the
discharge; reflection of light from objects surrounding the
experimental installation; a limited resource of the capacitor.

The statistic error $\sigma_{st}$ is caused by the following reasons:
the random character of geometric pattern of currents (see
Fig.1), the inaccuracy of setting the angles $\gamma$ and $\beta$; exactness of
instrumentation (power unit {\bf GOR-100}, voltmeter etc.); inaccuracies
of thermal detector and of constructions in Figs.5, 6,
non-controllable overheat of contacts; accidental changes in
exterior conditions (convective flows in the room, lighting
fluctuations, electromagnetic background).

Consider the systematic errors.  In the early experiments, the systematic
error caused by a prior warm-up of the plasma generator was
included into the final result. As was further clarified, there
were necessary on the average 7-10 shots before start of the
measurements (rotation of the plasma generator in space) to
reach operating conditions with a minimum error. If this error
taken into account, the heights of crosshatched triangles in
sectors 17-20 (Fig.5) can be reduced by $25\%$ making the result
more pronounced (in Fig.5 these errors are not accounted).  The
influence of build up of the cathode in the course of the
experiments at the rate of about 1mm per 30-40 shots on the
deflection of the oscillograph beam is not yet clear. This
factor led as to deterioration of sensitivity of measurements
(decrease in average deflection $L_{av}$ of the beam) so to an
increase of sensitivity. In the vernal and autumnal runs of
experiments, the said error was reduced to a minimum by way of
returning the plasma generator to its initial condition
(scraping bright the cathode and anode before each experiment).
In the course of experiment, the plasma generator was not
touched. The change of $L_{av}$ due to build-up of the cathode did
not exceed $2.5\%$ ($\sigma_k$).

In the long-term run of experiments
performed from 15.12.1999 till 3.05.2000, the burn-out of the
isolator led to a decrease of the total sensitivity of the
experimental technique, i.e. to some stable drift at a level of
$1\div2\%$ per experiment ($\sim 37$ shots). This drift was taken into
account and did not tell on the results of each separate
experiment so as on the final results, too, because the heights
of cross-hatched triangles in Fig.5 are given in percentage.

Before the new run of experiments with scanning the celestial
sphere, the plasma generator was replaced by a new one with the
same parameters. The grows of turbidity of the glass of the
thermal detector was resulting in a drift of the beam deflection
of the oscillograph and, hence, in a deterioration of total
sensitivity of the measuring system, but cleaning the glass 1-2
hours before each experiment weakened this effect down to a
level of $2\%$ which practically did not influence on the relative
value of the amplitude of bursts analyzed. Withdrawal of the
axis of the thermal detector from the direction of a maximum
luminous emittance of the discharge could led only to a drift
(or leaps) of $L_{av}$ no more than $1\%$ ($\sigma_m$).

The experiments with
the plasma generator were carried out in a windowless
underground room of the Physical Department of the Moscow State
University. The convective flows were practically absent, the
surround objects were always at the same places, there were no
electromagnetic noise from exterior sources.  In total  $\sim 1800$
shots were made in the course of experiments. The resource of
the new capacitor was about 10000 shots, therefore its
instability could not tell on the experimental results. The
summary systematic error ($\sigma_{syst}$) due to incontrollable processes
during the experiment is representable in the form
$$\sigma_{syst}=\sqrt{\sigma^2_k+\sigma^2_m}=\pm2,7\% $$
since those processes were independent from each other.

The statistic error caused by the random character of geometrical
pattern of discharge currents (see Fig.1) entered into the
summary error of experiment and was not determined separately.
For the experiments carried out from 15.12.1999 till 3.05.2000
(with rotating the plasma generator in the horizontal plane), the
summary error comprising the systematic and random ones is shown
in Table 2.

The accuracy of setting angles $\gamma$ and $\beta$ in the
course of experiments was no lesser than $0,5^\circ$ ($\sigma_o < 0,5\%$).
The precision of the power {\bf GOR-100} and the voltmeter equaled $0,5\%$
($\sigma_p$), that of the mirror-galvanometer oscillograph $\sim 2\%$ ($\sigma_o$).
The error of calibration of the thermal detector was no more
than $2\%$ ($\sigma_c$) (according to preliminary measurements). The
accuracy of construction was $\sim 1^\circ$.  An incontrollable overheat
of contacts was impossible in the experiments considered. Random
changes in luminous emittance of surround objects were absent.
According to the data presented, the total statistic error
($\sigma_{st}$) was no more than $\sim 2,8\%$. The total computation error in the
experiments was $\sim 3,9\%$. As is seen, it laid near to the
root-mean-square errors indicated in Tables 1,2 which enhanced
the validity of the results obtained. As can be seen from the
tables, in many experiments the amplitude of deflection of the
beam exceeded the root-mean-square error more than two times,
this also strengthens the plausibility of results.

\section{Conclusion.}

Thus it is shown in the present work that the spatial
distribution of the intensity of plasma radiation of pulsed
plasma generator is clearly anisotropic. A cone of directions is
observed in which plasma radiation reaches its maximum.  The
results of the experiments can be satisfactorily explained
basing on a hypothesis about the existence of cosmological
vectorial potential, a new fundamental vectorial constant
determining a new anisotropic interaction of objects in nature.

\section{Acknowledgments.}

The authors are grateful to participants of seminars held in Moscow
State University named by M.V.Lomonosov and in IOFRAN, as well
as personally to prof. A.A.Rukhadze for fruitful discussion of
results of investigation, to academicians of RAS S.T.Belyaev, V.A.Matveev,
V.M.Lobashev for the discussion of the connection of the result obtained
with the action of the new force manifesting itself in the $\beta$-decay
\cite{9}.

The authors give thanks also to the A.V.Chernikov for help in
performing the experiments, E.P.Morozov, L.I.Kazinova, and A.Yu.Baurov
for preparing the paper to print.

\pagebreak
\begin{mediumtext}

\begin{tabular}{|c|c|c|c|c|c|c|c|c|c|c|c|c|c|}
\hline
$\beta\backslash\gamma$&$270^\circ$&$260^\circ$&$250^\circ$&$240^\circ$&$230^\circ$&$220^\circ$&$210^\circ$&$200^\circ$&$190^\circ$&$180^\circ$&$170^\circ$&$L_{av}[mm]$&$\sigma\%$ \\
\hline
$-20^\circ$&70&70,5*&69,5&67&67,5&65&66&66&67&65&70&67,6&2,9\\
\hline
$-10^\circ$&69,5*&66&62,5&66,5&63&63&67&66&63&66,5&62,5&65,1&3,5\\
\hline
$0^\circ$&65&64&61&67,5&69*&66,5&65&60,5&63&65,5&66&64,8&3,8\\
\hline
$+10^\circ$&63,5&65&67*&63&65&62,5&66,5&61&63&63&63&63,9&2,7\\
\hline
$+20^\circ$&65&65&65&65,5&64,5&67&66,5&63,5&63,5&69*&63,5&65,3&2,5\\
\hline
$+30^\circ$&64&64,5&66&63&63,5&72,5*&69,5&59,5&66,5&72*&68&66,3&5,6\\
\hline
$+40^\circ$&71*&65&63&67,5&67,5&65,5&62&65,5&65&69,5&66&66,1&3,8\\
\hline
\end{tabular}
\end{mediumtext}
\vskip15pt
Table 1.
Deflection L of the oscillograph beam in the experiment of 03.11.2000,\\
$17^{22}\div19^{00}$ by Moscow time.
\vskip20pt
\begin{narrowtext}
\begin{tabular}{|c|c|c|c|c|}
\hline
N&Date&Time&Amplitude&Error\\
\hline
1&  15.12.1999&  $16^{40}-17^{15}$& 25,7 \% &  $\pm 3,7$ \%\\
\hline
 2&15.12.1999&$17^{20}-17^{50}$& 11,8 \% & $\pm 3,6 $\%\\
\hline
 3&20.01.2000& $15^{45}-16^{10}$& 24 \% &   $\pm 3,3 $\%\\
\hline
 4&20.01.2000&$  16^{20}-16^{40} $& 3,5 \% &  $\pm 3 $\%\\
\hline
 5&21.01.2000&$14^{45}-15^{15}  $ & 7 \%  &                       $\pm 4 $\%      \\
\hline
 6&21.01.2000&$15^{30}-16^{40}  $&  22 \% &                       $\pm 5,2 $\%    \\
\hline
 7&2.02.2000 & $14^{00}-14^{40}$&16,5 \%&  $\pm 5,2 $\%           \\
\hline
 8&2.02.2000  &$14^{55}-15^{30}    $&17,9 \%&  $\pm 4,5 $\%    \\
\hline
 9&9.02.2000  &$13^{25}-14^{00}     $&11,3 \%&  $\pm 6 $\%            \\
\hline
 10&9.02.2000 &$14^{10}-14^{40}      $& 10 \% &   $\pm 4,7 $\%   \\
\hline
 11&16.02.2000 &$12^{55}-13^{20}       $&11,8 \%&  $\pm 4,8 $\%         \\
\hline
 12&16.02.2000 &$13^{42}-14^{10}        $&8,7 \% &  $\pm 4 $\%                 \\
\hline
 13&23.02.2000&$13^{00}-13^{55}         $& 19,8 \%&  $\pm 5,1 $\%   \\
\hline
 14&23.02.2000&$14^{00}-14^{45}          $ &11,6 \%&  $\pm 4,2 $\%         \\
\hline
 15&1.03.2000&$11^{50}-12^{40}           $ & 11 \%  &  $\pm 4,5 $\%               \\
\hline
 16&1.03.2000&$12^{45}-13^{20}            $ & 12 \%  &  $\pm 5 $\%          \\
\hline
 17& 9.03.2000&$11^{20}-12^{07}           $ & 7,8 \%&   $\pm 2,6 $\%    \\
\hline
 18& 9.03.2000&$12^{20}-12^{58} $  &13,9 \%&  $\pm 4,6 $\%          \\
\hline
 19& 15.03.2000&$10^{45}-11^{30}  $  &5 \%  &   $\pm 2,5 $\%   \\
\hline
  20&15.03.2000&$11^{40}-12^{30} $& 11,3 \%&  $\pm 3 $\%       \\
\hline
  21&22.03.2000&$10^{15}-11^{07}  $ &9,3 \% &  $\pm 3 $\%       \\
\hline
  22&22.03.2000&$11^{15}-12^{00}   $ & 10,2 \% & $\pm 3,6 $\%       \\
\hline
  23&29.03.2000&$10^{45}-11^{55}    $ &8 \%   &  $\pm 4 $\%    \\
\hline
  24&29.03.2000&$12^{00}-12^{35}     $ &7,6 \% &  $\pm 3,5 $\%   \\
\hline
  25&5.04.2000&$10^{25}-11^{10}      $ & 7 \%  &  $\pm 3 $\%   \\
\hline
  26&5.04.2000&$11^{15}-11^{55}       $ & 8,6 \% &  $\pm 5 $\% \\
\hline
  27&12.04.2000&$11^{00}-11^{40}        $ &12,2 \%&  $\pm 3 $\%       \\
\hline
  28&12.04.2000&$11^{45}-12^{22}         $ &9,5 \% &  $\pm 5 $\%   \\
\hline
  29&26.04.2000&$10^{17}-11^{00}          $ &4,5 \% &  $\pm 2,6 $\%  \\
\hline
  30&26.04.2000&$11^{05}-12^{45}           $ &4,6 \% &  $\pm 3 $\%   \\
\hline
  31&3.05.2000& $11^{07}-11^{52}$  &6,2 \% &  $\pm 2 $\% \\
\hline
  32&3.05.2000&$12^{25}-13^{07}$  &4 \%    & $\pm 2 $\%\\
\hline
\end{tabular}
\end{narrowtext}
\vskip15pt
Table 2.
{\it Amplitudes} of maximum deflection (in percent from the average) with indication of
date, time (Moscow), and root-mean-square {\it error} of the experiments.
% N   Data and timeAmplitude ofmaximumdeflection ofthe beamRoot-mean-square errorof  theexperiment
\pagebreak
\vspace*{5cm}

%Subscriptions of the figures:
%
%Fig. 1. The diagram of the measuring device.
%Fig. 2.  The volt-ampere characteristics of the discharge of the plasma generator
%	I - current, U - voltage.
%Fig. 4. The magnitude L (mm) of beam deflection of the mirror-galvanometer oscillograph in dependence
%of angle of rotation $\theta$ for the experiments of  15.12.1999 ($16^{40}-17^{15}$) ($\Box$) and 20.01.2000
%($17^{20}-17^{50}$) ($\bullet$).
%
%Fig.5. Directions of the axis of the plasma generator along which maximum deflections of the oscillograph
%beam (indicated by arrows) when rotating the plasma generator in the horizontal plane, were observed.
%By H start of rotation ($\theta = 0$) is denoted.
%Indicated are data and Moscow times of observation of maximums in beam deflection.
%At the center of the Figure, heights of cross-hatched triangles correspond to the sums of beam
%deflection magnitudes (in percent) from their average value exceeding the error (for a given sector).
%${\bf A}_g$  is the cosmological vectorial potential..
%
%Fig.6. The relative spatial coordinates and directions of the axis of the plasma generator corresponding to
%maximum deflections of the beam of the oscillograph in the experiments carried out 10.05.2000
%till 31.05.2000 (denoted by circles $\circ$) and from 11.10.2000 till 3.11.2000 (denoted by asterisks *).
%The vernal experiments are related to $11^{25}$ by Moscow time of 31.05.2000, and the autumnal
%experiments are to $20^{00}$ of 11.10.2000 by the same time.
%${\bf A}_g$ is direction of the cosmological vectorial potential.

\begin{figure}
\epsfxsize=8,5cm\epsfbox{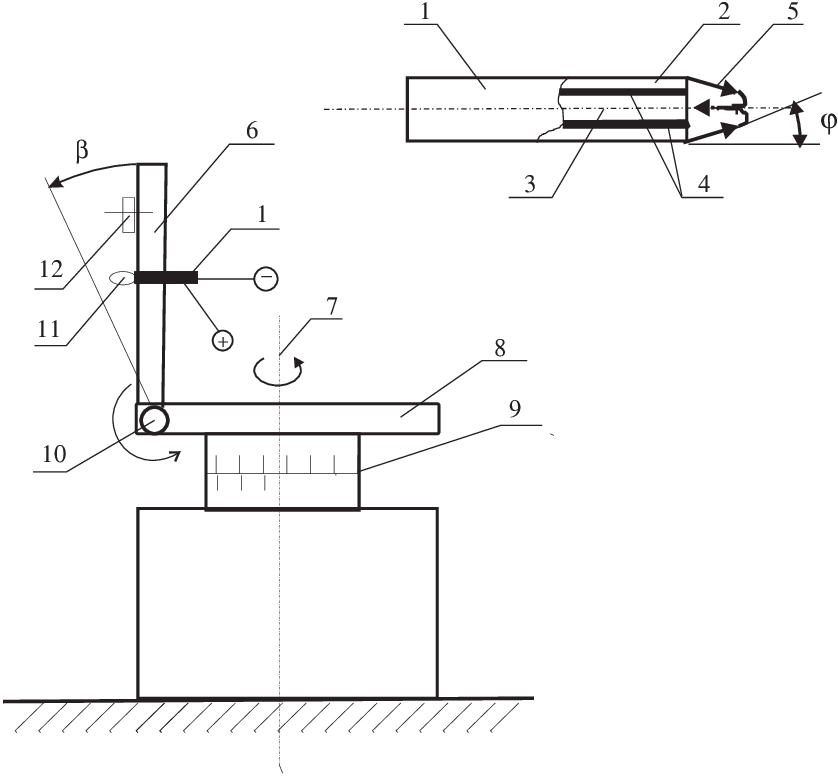}
\caption{The diagram of the measuring device.}
\end{figure}
\begin{figure}
\epsfxsize=8,5cm\epsfbox{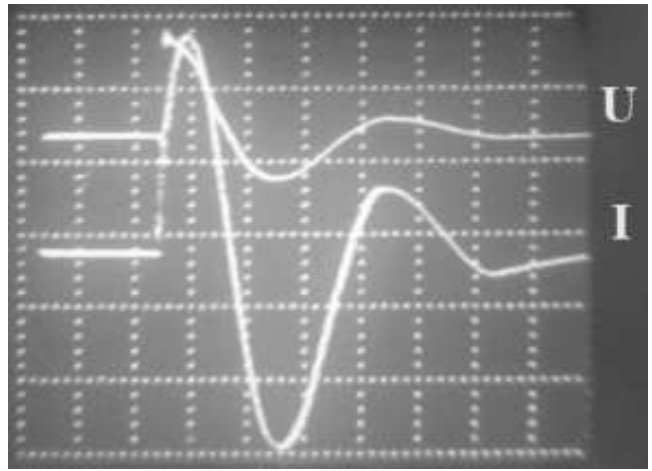}
\vskip5pt
\caption{The volt-ampere characteristics of the discharge of the plasma generator
I - current, U - voltage.}
\end{figure}
\begin{figure}
\epsfxsize=8,5cm\epsfbox{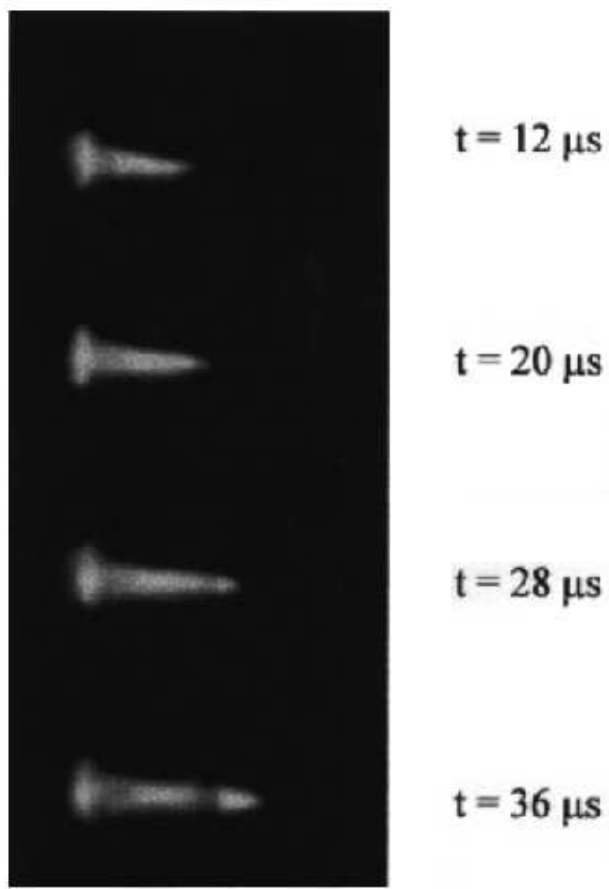}
\caption{A fragment of plasma jet rapid shooting film.}
\end{figure}
\vspace*{-1cm}
\begin{figure}
\epsfbox{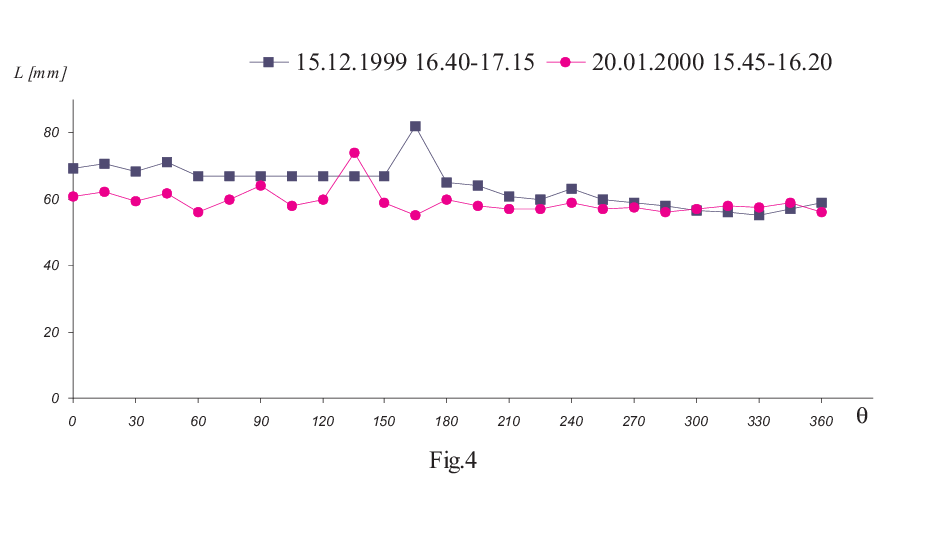}
\widetext
%\vspace*{-2cm}
\caption{The magnitude L (mm) of beam deflection of the mirror-galvanometer oscillograph in dependence
of angle of rotation $\theta$ for the experiments of  15.12.1999 ($16^{40}-17^{15}$) ($\Box$) and 20.01.2000
($17^{20}-17^{50}$) ($\bullet$).}
\end{figure}
\begin{onecolumn}
\begin{figure}
\epsfxsize=17cm\epsfbox{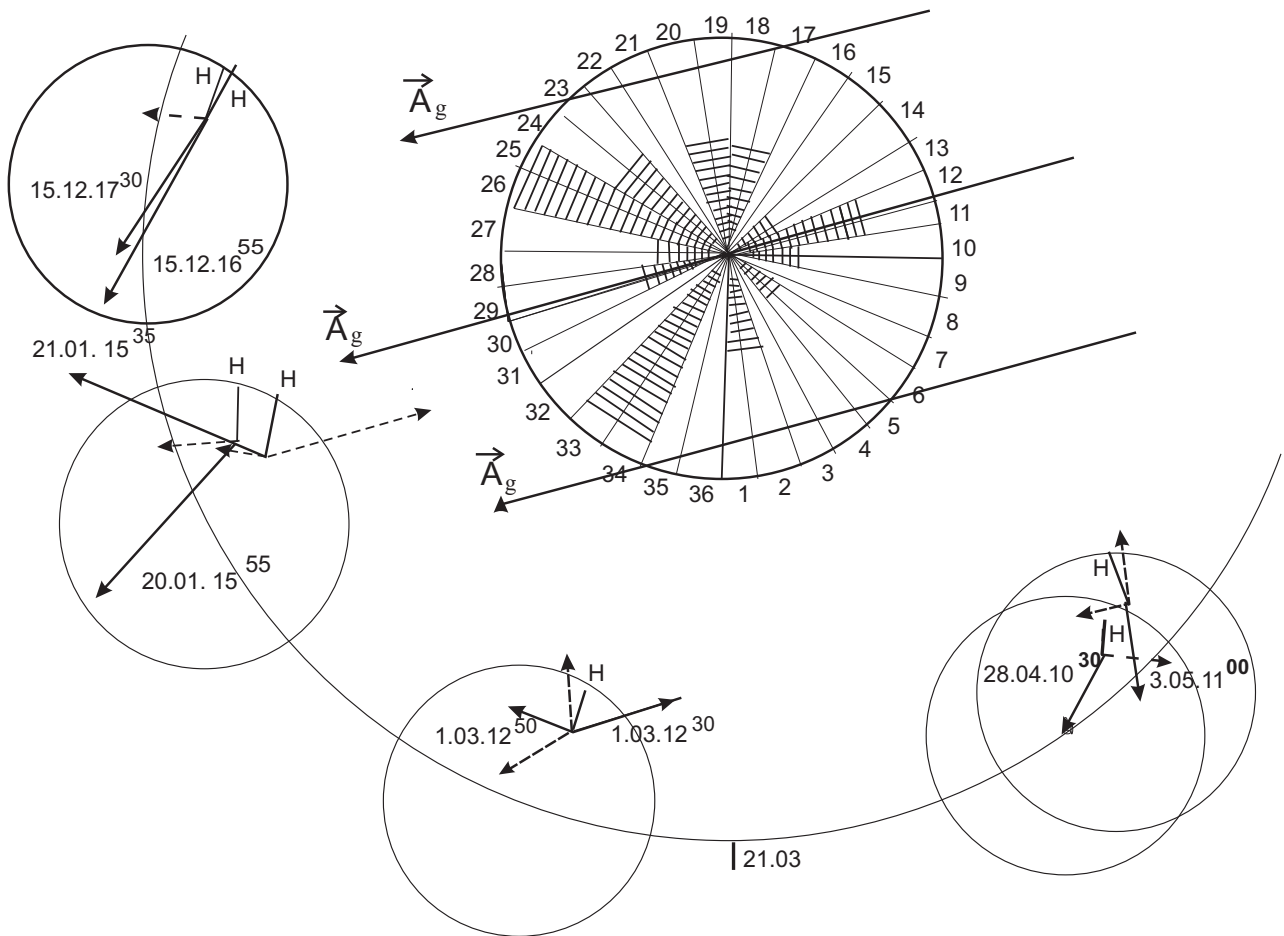}
\caption{
Directions of the axis of the plasma generator along which maximum deflections of the oscillograph
beam (indicated by arrows) when rotating the plasma generator in the horizontal plane, were observed.
By H start of rotation ($\theta = 0$) is denoted.
Indicated are data and Moscow times of observation of maximums in beam deflection.
At the center of the Figure, heights of cross-hatched triangles correspond to the sums of beam
deflection magnitudes (in percent) from their average value exceeding the error (for a given sector).
${\bf A}_g$  is the cosmological vectorial potential.}
\end{figure}
\begin{figure}
\epsfbox{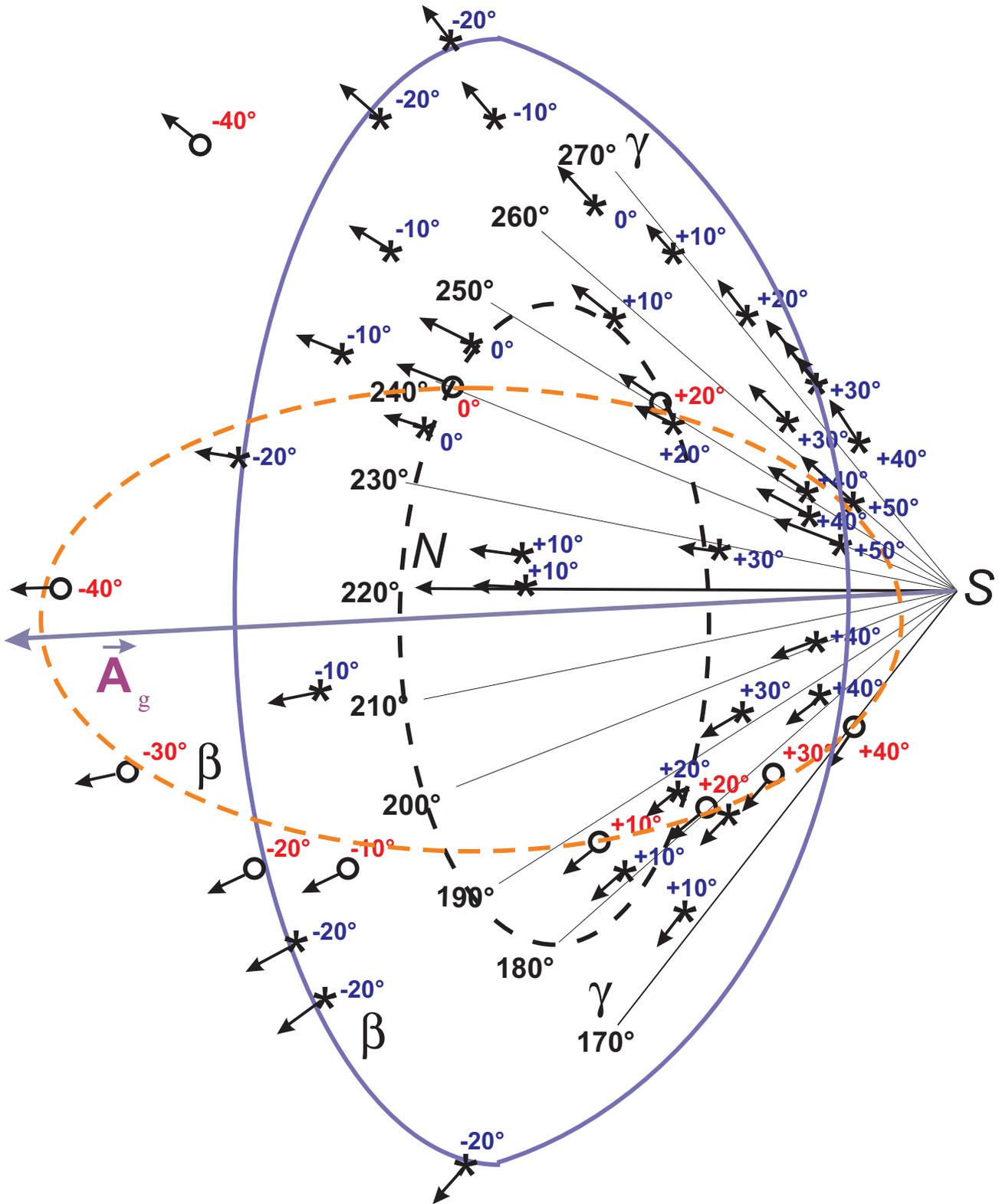}
\caption{The relative spatial coordinates and directions of the axis of the plasma generator corresponding to
maximum deflections of the beam of the oscillograph in the experiments carried out 10.05.2000
till 31.05.2000 (denoted by circles $\circ$) and from 11.10.2000 till 3.11.2000 (denoted by asterisks *).
The vernal experiments are related to $11^{25}$ by Moscow time of 31.05.2000, and the autumnal
experiments are to $20^{00}$ of 11.10.2000 by the same time.
${\bf A}_g$ is direction of the cosmological vectorial potential.}
\end{figure}
\end{onecolumn}

\end{document}